\documentclass[preprint,12pt, a4paper]{elsarticle}

\usepackage{amsmath}
\usepackage{graphicx}
\usepackage{dcolumn}
\usepackage{array}
\usepackage{bm}
\usepackage{fancyvrb}
\usepackage{bera}
\usepackage{fixltx2e}
\usepackage{float}
\restylefloat{table}

\journal{SoftwareX}

\begin{document}
\bibliographystyle{unsrt}
\begin{frontmatter}

\title{OPENMMF: a library for multimode driven quantum systems}

\author{German A. Sinuco-Leon}
\address{School of Mathematics and Physical Sciences, University of Sussex, Falmer, BN1 9QH, United Kingdom.}

\begin{abstract}
OPENMMF is a numerical library designed to evaluate the dynamics of quantum systems with a discrete spectrum and driven by an arbitrary combination of harmonic couplings. The time-evolution operator of such systems is calculated as a multifrequency Fourier expansion, which results from expressing the time-dependent Schr\"{o}dinger equation in the frequency domain \cite{ho1983semiclassical}. The library provides a generic tool to study systems with arbitrary spectral composition, limited only by the available computational resources. OPENMMF includes functionalities to build dense and sparse matrix representations of the Hamiltonian and functions to calculate the micromotion operator and time/phase average of state populations. The library introduces a generalised notion of \textit{dressed state} for systems with polychromatic driving.  In this paper, we describe the design and functionality of OPENMMF, provide examples of its use and discuss its range of applicability in problems of current interest in quantum dynamics. The library is written in object-oriented style Fortran90 and includes a set of wrappers for C++ and Python..
\end{abstract}

\begin{keyword}
Polychromatic driving \sep Multimode Floquet \sep Quantum dynamics \sep Fortran \sep C++ \sep Python \sep Dressed states \sep non-equilibrium quantum dynamics
\end{keyword}

\end{frontmatter}

\section*{Required Metadata}
\label{}

\section*{Current code version}
\label{}


\begin{table}[H]
\begin{tabular}{lp{9.5cm}}
\textit{Manuscript Title}: & OPENMMF: A library for multimode driven quantum systems \\
\textit{Authors}: & G. A. Sinuco-Leon\\
\textit{Program Title}:& OPENMMF\\
\textit{Licensing provisions}:& CC BY 4.0 \\
\textit{Programming language}:& Fortran 90, C++, python\\
\textit{Operating system}: & Linux, Mac, Cygwin\\
\textit{RAM}:& size-dependent\\
\textit{Number of processors used}:& single, user-configurable\\
\textit{Keywords}: &Polychromatic driving, Multimode Floquet, Quantum dynamics, Dressed states \\
\textit{Subprograms used}:& LAPACK, MKL-intel, python: numpy, ctypes \\
\textit{Nature of problem}:& Numerical solution of the Schrodinger equation with harmonic time dependence\\
\textit{Solution method}:& Diagonalisaton of the multifrequency Floquet matrix. \\
\textit{Source code repository}:& https://github.com/OPENMMF
\end{tabular} 
\end{table}


\section{\label{sec:motivation} Motivation and significance}

The dynamics of quantum systems driven by polychromatic (or multimode) electromagnetic radiation is of interest to many domains of physical sciences. Over the last few decades, there has been a rapid development of experimental platforms where coherent dynamics of driven quantum systems has been explored with unprecedented precision and sensitivity. Examples of this include experiments with ultracold atomic ensembles \cite{nakajima2016topological,PhysRevA.97.013616}, trapped ions \cite{PhysRevLett.117.220501}, Nitrogen-Vacancy centres \cite{balasubramanian2009ultralong,golter2014protecting,zhou2017holonomic} and superconducting circuits \cite{tuorila2010stark,yan2013rotating,vion2002manipulating}. In all these platforms, the state of an ensemble of quantum systems is manipulated by applying a combination of monochromatic driving fields, produced by lasers or electric/electronic devices. Technical advances of such controlling techniques have led to practical applications (e.g. high-precision spectroscopy \cite{baumgart2016ultrasensitive} or Nuclear Magnetic Resonance \cite{vandersypen2005nmr}), and are instrumental to developing quantum technologies for sensing and computing \cite{o2009photonic}.

A growing number of applications exploit the possibility to modify the properties of a quantum system through its interaction with electromagnetic radiation. The phenomena of electromagnetic induced transparency (EIT) is a textbook example of this situation \cite{scully1999quantum}: The absorption spectrum of a three-level system is strongly modulated by coupling them through a monochromatic field. Expanding this idea to more complex quantum systems has lead to applications such as robust atomic clocks \cite{sarkany2014controlling,kazakov2015magic}, ultra-sensitive field sensors with NV-centres \cite{baumgart2016ultrasensitive}, and to discovering new physical phenomena, e.g. topological features in driven many-body systems \cite{lindner2011floquet} and time crystals \cite{zhang2017observation}, which can be implemented with solid state platforms \cite{mukherjee2020observation} and ultracold-atomic ensembles loaded into optical lattices \cite{gross2017quantum}.

\begin{figure}
\centering
\includegraphics[width=0.9\linewidth]{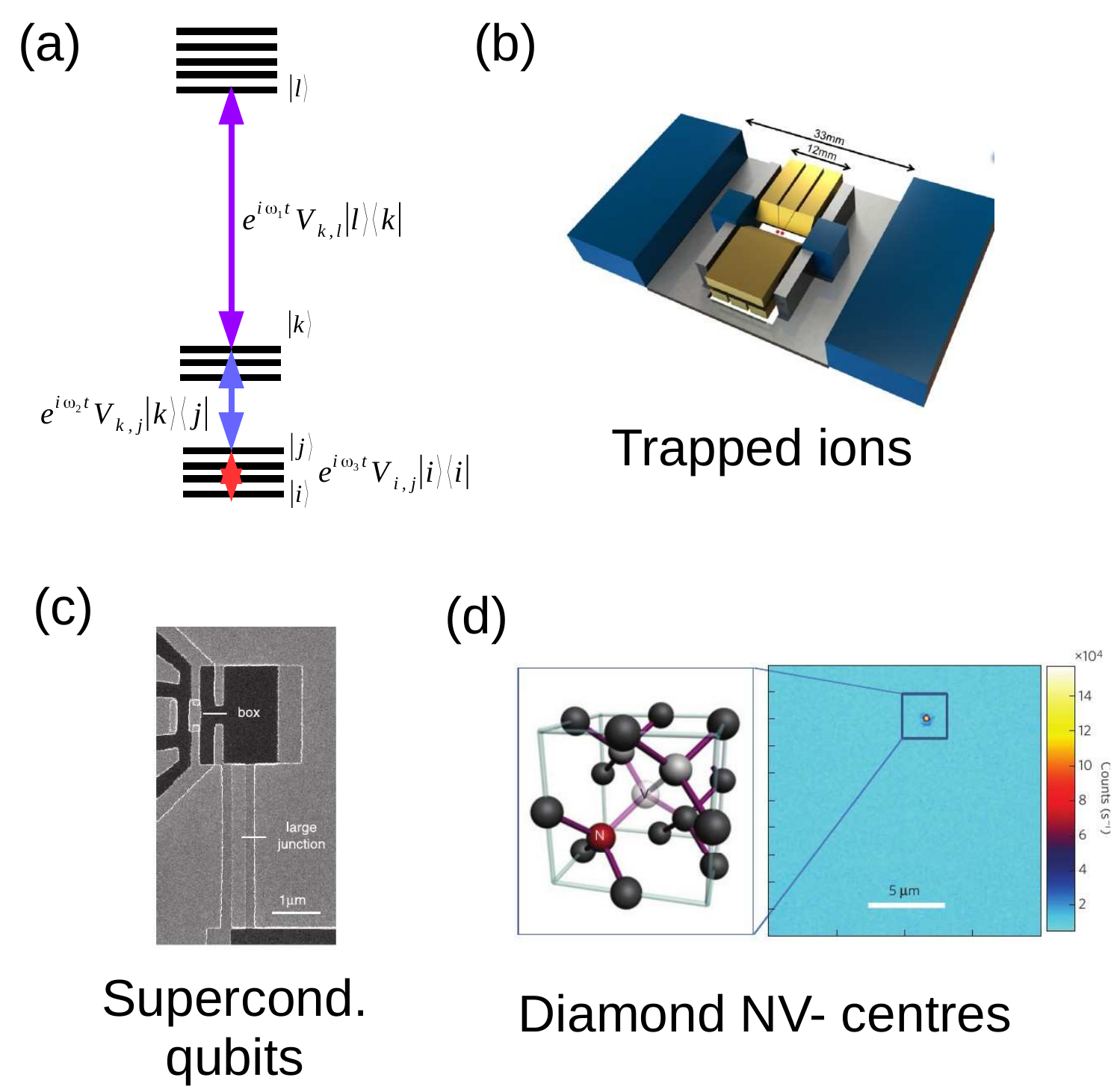}
\caption{\label{fig:SystemSketch} (a) Schematic energy level structure of a generic quantum system. The basis of states consists of a discrete set of energy states, which define several bands according to the level energy spacing. Inter- and intra-band coupling is induced by electromagnetic radiation tuned at the corresponding frequencies, as indicated by the coupling terms. The wide variety of physical systems described by this model includes (b) trapped ions \cite{PhysRevLett.117.220501}, (c) superconducting qubits \cite{vion2002manipulating} and (d) diamond NV-centres \cite{balasubramanian2009ultralong}.}
\end{figure}

As schematically shown in Fig. \ref{fig:SystemSketch}, the physical systems mentioned above have a similar structure: they are described by a discrete set of states $\{\left| i\right\rangle\}$ and polychromatic (or multimode) couplings. The OPENMMF library, introduced in this work, provides tools to evaluate the time-evolution operator of such driven systems. The library implements the multimode Floquet method described originally by Ho, Chu and Tiet in \cite{ho1983semiclassical}, where the semiclassical time-dependent Schr\"{o}dinger equation is substituted by a time-independent matrix representation in the frequency-domain. The library also provides functions to evaluate physical quantities that cannot be easily accessed from direct integration of the Schr\"{o}dinger equation in the time-domain, such as the micromotion operator and dressed energies and states \cite{shirley1965solution}. 

The OPENMMF library includes algorithms that enable its use for a wide variety of problems currently relevant in quantum physics. For example, in the context of quantum technologies, the library can be used to design sequences of electromagnetic pulses for qubit gates \cite{song2016fast,zhou2017holonomic} as well as elements for manipulating matter-waves \cite{weidner2018experimental}. The library is also ready to assist emerging areas of research such time-domain quantum simulations \cite{price2015four,martin2017topological,delande2017three} and (multimode) Floquet engineering \cite{goldman2016topological}, where harmonic driving of the system plays an essential role.


\section{\label{sec:FloquetBloch} Time-evolution of multimode driven quantum systems}

The main task of the OPENMMF library is to calculate the time-evolution operator, $U(t',t), ~ t'>t$, of systems whose Hamiltonian has the form:
\begin{equation}
H = \sum_{i,j}^D E_{i,j} \left| i\right\rangle \left\langle j \right| + \sum_{i,j}^D \sum_{\ell=1}^N \sum_{n \in Z} V_{i,j}^{\ell,n} e^{i n \omega_\ell t} \left| i\right\rangle \left\langle j \right| + \textrm{h.c.}
\label{eq:Hamiltonian}
\end{equation}
where $D$ is the dimension of the Hilbert space, ${E_{i,j}}$ defines the static component of the Hamiltonian $H$, $V_{i,j}^{\ell,n}$ is the coupling between the states $i$ and $j$ oscillating at frequency $n \omega_{\ell}$ (i.e. the $n$-th harmonic of the $\ell$-th fundamental frequency $\omega_{\ell}$) and $N$ is the number of non-commensurable frequencies.

To calculate the time-evolution operator, we generalise the Rotating (or Resonant) Wave Approximation (RWA) taking into account the complex time dependence of Eq.~ (\ref{eq:Hamiltonian}). For this, we rephrase the problem in terms of building a time-dependent unitary transformation, $U_F(t)$ to a new basis $\{\left| \bar{i} \right\rangle\}$, that leads to a \textit{time-independent} and diagonal Hamiltonian, $\bar{H}$. The standard quantum-mechanical transformation rule to the Schr\"{o}dinger equation \cite{chu1985recent,PhysRevA.81.063626} leads to the condition:
\begin{eqnarray}
 U_F^\dagger(t) \left[ H(t) - i \hbar \partial_t \right] U_F(t)  &=& \sum_{\bar{i}} \bar{E}_{\bar{i}} \left| \bar{i} \right\rangle \left\langle \bar{i} \right|
\label{eq:Hdressed}
\end{eqnarray}

In the basis of states defined by this transformation, $\{\left|\bar{i}\right\rangle\}$. the time evolution operator is diagonal and has the form:
\begin{equation}
\bar{U}(t',t) = \sum_{\bar{i}} e^{-i \bar{E}_{\bar{i}} (t'-t)} \left| \bar{i} \right\rangle \left\langle \bar{i} \right|
\label{eq:dressedtimeevolution}
\end{equation}
which let us to calculate the time evolution operator in the original basis, $\left\{ \left| i\right\rangle\right\}$, just by inverting the transformation $U_F(t)$ according to \cite{PhysRevA.81.063626}:
\begin{equation}
U(t',t) = U_F(t') \bar{U}(t',t) U_F(t)
\label{eq:baretimeevolution}
\end{equation}

To formulate a fully defined computational problem, we express the micromotion operator $U_F(t)$ as the multifrequency Fourier series \cite{ho1983semiclassical}:
\begin{equation}
U_F(t) = \sum_{\vec{n}} U_{i,\bar{i}}^{\vec{n}} e^{-i\vec{\omega} \cdot \vec{n}t} \left| i \right\rangle \left\langle \bar{i} \right|
\label{eq:micromotionexpansion}
\end{equation}
where $\vec{\omega} = (\omega_1,\omega_2,\ldots,\omega_N)$ and $\vec{n}$ is a $N$-dimensional vector of integers. Using this expansion in Eq.~(\ref{eq:Hdressed}) and performing an integral over time, we obtain a fully defined eigenproblem for the eigenvalues $\bar{E}_{\bar{i}}$ and Fourier components of the unitary transformation $U_{i,\bar{E}}^{\vec{n}}$:
\begin{equation}
\sum_j(E_{i,j} - \hbar \vec{n} \cdot \vec{\omega})U^{\vec{n}}_{j,\bar{i}} + \sum_{j} \sum_{\vec{m}} \left[ V^{\vec{m}}_{i,j} U^{\vec{n}+\vec{m}}_{j,\bar{i}} + V^{\vec{m}*}_{j,i} U^{\vec{m}-\vec{n}}_{j,\bar{i}}\right] = \bar{E}_{\bar{i}}U^{\vec{n}}_{i,\bar{i}}
\label{eq:multimodeeigenproblem}
\end{equation}
where the couplings $V_{i,j}^{\ell,n}$ define the values of $V_{i,j}^{\vec{n}}$ with the vector $\vec{m} = (0,\ldots , m, \ldots, 0)$, where the value $m$ locates at the $\ell-$th position. To obtain a finite matrix representation of this problem we truncate the sum over the number of modes of the Fourier expansion Eq.~ (\ref{eq:micromotionexpansion}), which corresponds to the number of Floquet manifolds for each fundamental frequency. In Appendix A, we show an specific example of the shape of the matrix for a bichromatic driven problem. 

This formulation to calculate the time-evolution operator is equivalent to the multimode Floquet representation of the Hamiltonian that introduces the extended Hilbert space $\left| E_i,\vec{n} \right\rangle$  \cite{ho1983semiclassical,shirley1965solution,verdeny2016quasi}. However, the semiclassical description presented here makes emphasis in the experimentally accessible states, which are commonly used to express the static part of the Hamiltonian Eq.~  (\ref{eq:Hamiltonian}). 


\section{Software description}
\begin{figure}[!htb]
\centering
\includegraphics[width=0.9\linewidth]{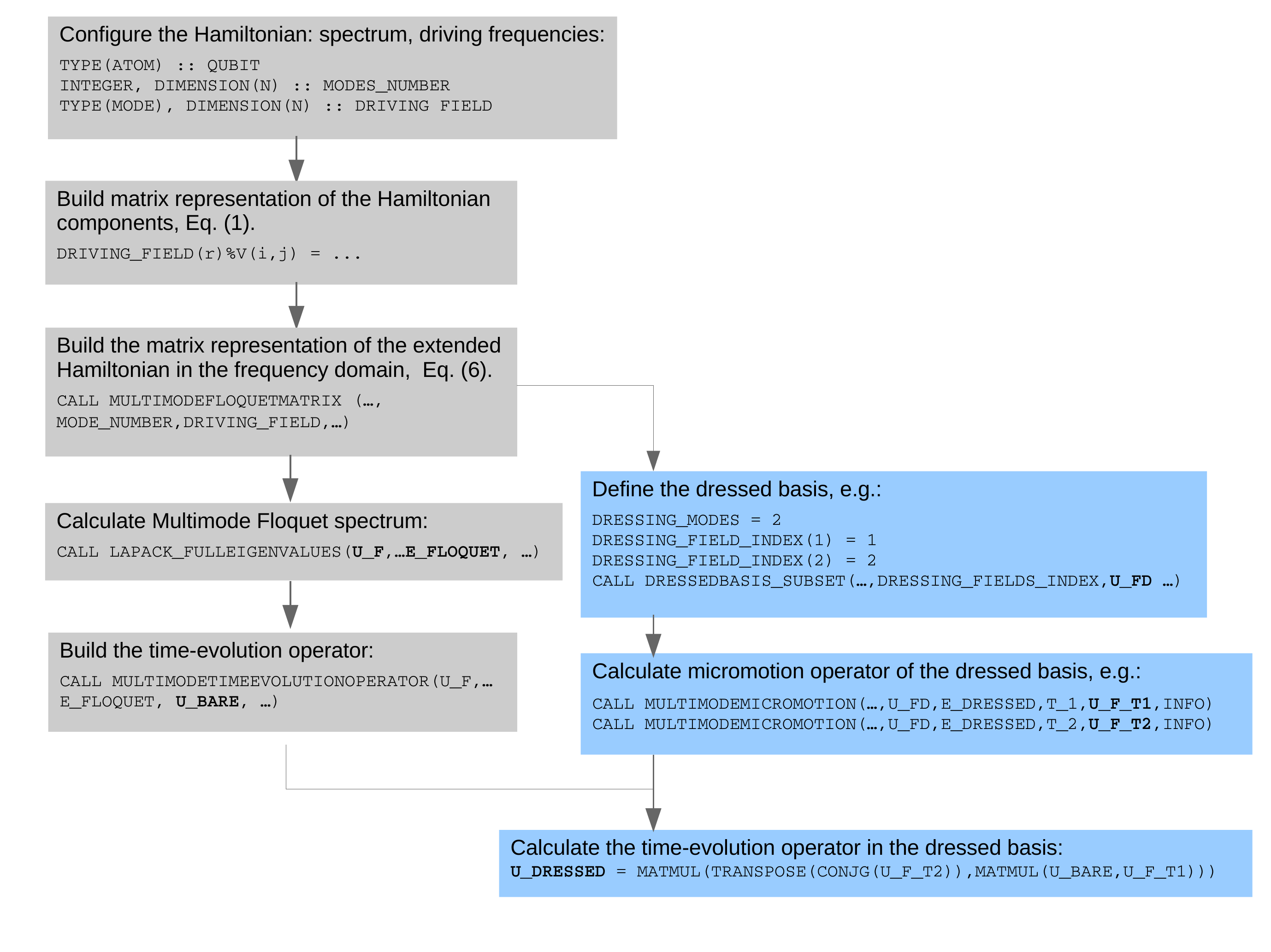}
\caption{\label{fig:FlowChart} Typical sequence of function calls to calculate the time-evolution operator of a quantum system with the OPENMMF library. After defining the parameters of the Hamiltonian Eq.~ (\ref{eq:Hamiltonian}), the user can build the matrix representation of the Schr\"odinger equation in the frequency space. The eigenvalues and eigenvectors of this matrix can then be used to evaluate the time-evolution operator. Alternatively, the user can select a subset of the driving fields to define a set of dressed states and evaluate the time-evolution operator in such a basis. Normal and bold fonts indicates input and output subroutine arguments, respectively.}
\end{figure}

The OPENMMF library is written in object-oriented style FORTRAN 90 and includes wrappers to enable its use with C++. The software architecture resembles a typical sequence of operations enabled by this library, which is shown in Fig. \ref{fig:FlowChart}. 

The starting point to use the OPENMMF library is defining parameters to build a matrix representation of the Hamiltonian components of Eq.~ (\ref{eq:Hamiltonian}). Then, the user can execute a series of tasks that lead to the time-evolution operator.

The components of the library can be classified into four groups:
\begin{enumerate}
\item Fortran modules that define as global variables relevant physical constants (\verb physical_constants ) and properties of atomic systems (\verb ATOMIC_PROPERTIES ). 
\item Two Fortran derived data types (and corresponding C++ structures) that define the parameters required to build a matrix representation of the system's Hamiltonian (\verb TYPE::ATOM  and \verb TYPE::MODE ).
\item A set of computational routines that perform distinct computational tasks, as in Fig. \ref{fig:FlowChart}.
\item A set of driver routines, each one to call a sequence of computational routines for solving standard problems with a single subroutine call.
\end{enumerate}

To give a brief overview of the library structure and functionalities, we describe the last three groups. We refer the reader to the user manual for detailed information about the Fortran modules defined in the library, usability of the C++ and Python wrappers and performance of the library.  \cite{openMMFUserManual} 


\subsection{Setting the system configuration}
The parameters required to build a matrix representation of the Hamiltonian Eq.~ (\ref{eq:Hamiltonian}) must be initialised using the following three data structures: 
\begin{enumerate}
\item The Fortran derived type \verb TYPE::ATOM , which contains one component to set the dimension of the Hilbert space (\verb D_Bare ) and another one (declared as an array of type \verb double ), which specifies the  energy spectrum of the system.
\item An array of \verb integer  type, with as many components as the number of fundamental  driving frequencies plus one (i.e $N+1$). The entries of this array indicate the number of harmonics of each driving frequency, with the first component corresponding to the static component of the field.
\item An array of the Fortran derived type \verb TYPE::MODE , with as many elements as the total number of driving frequencies plus one (i.e. $1 + \sum_{\ell=1}^N n_\ell $). The components of this derived data type include the driving angular frequency (\verb omega ), complex matrices to store the explicit representation of the coupling matrix ($V$), and the number of modes of the Fourier expansion \verb N_Floquet . 
\end{enumerate}


\subsection{\label{sec:sequence} Typical computing sequences}

Using the previous three data structures and an explicit representation of the components of the Hamiltonian,  the user can proceed to build a finite matrix representation of the eigenproblem Eq.~ (\ref{eq:multimodeeigenproblem}). This task can be done using the subroutine \verb MULTIMODEFLOQUETMATRIX , which creates a square-matrix representation of the right-hand-side of Eq.~ (\ref{eq:micromotionexpansion}).

The eigenvalues and eigenstates of the extended representation of the Hamiltonian can then be obtained using standard diagonalisation routines. The library includes wrapers to call the relevant subroutines of the LAPACK and the MKL-Intel libraries. However, the user is free to use other tools. The resulting eigenvalues and eigenvectors are then employed to calculate the time-evolution ($U(t',t)$, Eq.~(\ref{eq:dressedtimeevolution})) and micromotion ($U_F(t)$, Eq.~(\ref{eq:micromotionexpansion})) operators, using the routines \verb MULTIMODETIMEEVOLUTIONOPERATOR  and \verb MICROMOTIONOPERATOR , respectively.  

The openMMF library includes the driver subroutine \verb TIMEEVOLUTIONOPERATOR ,  which allows the user to perform these three tasks (building the Hamiltonian representation, diagonalisation and evaluation of $U(t',t)$) with a single subroutine call after defining the components of the Hamiltonian. 

To take advantage of the sparse nature of Eq.~(\ref{eq:multimodeeigenproblem}), the library also allows us to create a sparse representation of this equation and diagonalisation with the MKL-intel library. The names of the corresponding subroutines have appended the letters \verb _SP , and the user must ensure the correct allocation of memory.


\subsection{\label{sec:dressedbasis} Time-evolution in a dressed basis}
\begin{figure}[!htb]
\centering
\includegraphics[width=0.7\linewidth]{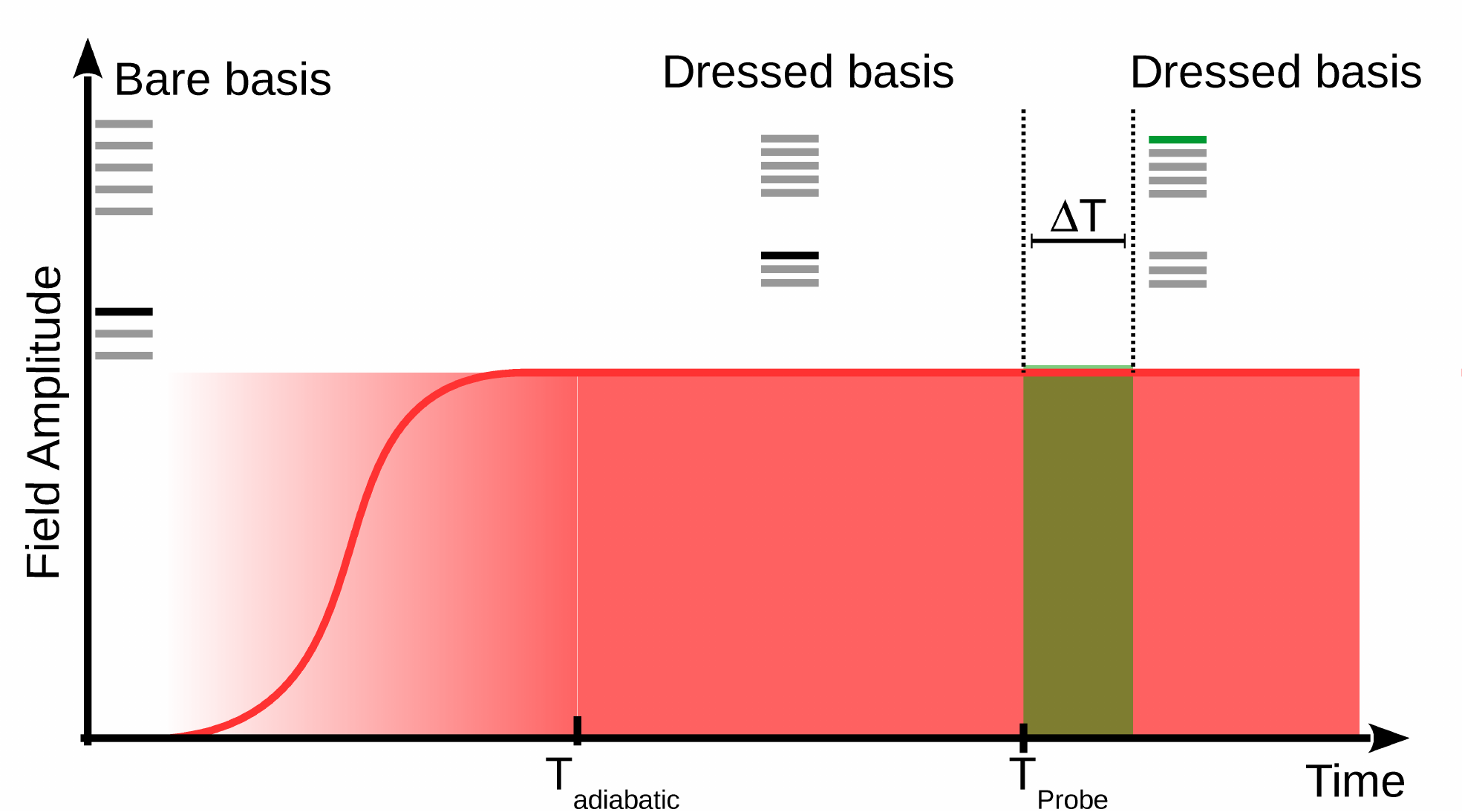}
\caption{\label{fig:DressedBasis} Typical time sequence to prepare and probe dressed states. The dressing modes are adiabatically switched on and kept constant during the probing period. }
\end{figure}

The openMMF library provides tools to evaluate the dynamics of driven quantum systems in the so-called \textit{dressed basis}. Figure \ref{fig:DressedBasis} presents a situation where it is convenient to express the time evolution operator on a time-dependent basis that differs from the one used to represent the Hamiltonian. 

In this case, the driving modes applied to the quantum system can be classified into two groups: dressing and probing modes. Initially, the dressing modes are adiabatically switched on over a period of time $T_{\textrm{adiabatic}}$, after which their amplitudes and frequencies remain constant. This procedure is shown schematically in Fig.~\ref{fig:DressedBasis}, with the red line and the colour gradient indicating the slow change of the amplitudes of the dressing modes. The adiabatic change of the dressing modes ensures that the basis of the static system (on the left-hand-side of the Fig.~\ref{fig:DressedBasis}) deforms into the time-dependent basis defined by the transformation $U_F(t)$, Eq.~(\ref{eq:multimodeeigenproblem}) defined by \textit{only} the dressing modes. Thus, for example, when at $T=0$ the system is prepared in the state with the $i$-th energy, $\left| i \right\rangle$, after $T_{\textrm{adiabatic}}$, this state becomes the corresponding linear superposition associated with the $\bar{i}$-th eigenvalue $\bar{E}_{\bar{i}}$. 

After a period of evolution driven by the dressing modes, the system is probed by pulses of the probing subset, while the dressing modes are kept constant. During the duration of the probing pulses, $t\in [T_{\textrm{Probe}},T_{\textrm{Probe}} + \Delta T]$, the time-evolution of the system is determined by a Hamiltonian of the form Eq.~ (\ref{eq:Hamiltonian}), which must include \textit{both} the dressing and probing modes.  The OPENMMF evaluates the time evolution operator initially in the bare basis and includes functions to define the micromotion operator associated with the subset of dressing fields and the corresponding transformation of the time-evolution operator. An example of this situation is shown in the following section.


\subsection{\label{sec:functionalities} Functionality for spin systems}

Spin systems are of particular interest due to the simplicity of their energy spectrum and their experimental feasibility with a diverse set of experimental platforms. For these systems, the couplings to harmonic drivings are of the form:
\begin{equation}
H^{\ell,n} = \sum_{i\in{x,y,s}} e^{n\omega_\ell t +\phi^{\ell,n}_i} V_i^{\ell,n} S_i
\label{eq:SpinHcomponents}
\end{equation}
where $S_i$ is the $i$-th Cartesian component of angular momentum operator, whose matrix representation follows the standard definition in Sec. 3.5 of \cite{sakurai1995modern}. Each one of these terms is defined by six real parameters: three  phases ($\phi^{\ell,n}$) and three amplitudes ($V_i^{\ell,n}$), which are included as components of the Fortran data structure \verb TYPE::MODE . The OPENMMF library includes the subroutine \verb SETHAMILTONIANCOMPONETS  that uses these parameters to define Hamiltonian components Eq.~ (\ref{eq:SpinHcomponents}).

The library can also handle the energy configuration of systems composed by two spins $S_1$ and $S_2$, coupled via the dipole interaction $H_{\textrm{dipole}} = A \boldsymbol{S_1} \cdot \boldsymbol{S_2}$ and each spin interactions as above in Eq.~(\ref{eq:SpinHcomponents}). The library evaluates the Hamiltonian and time-evolution operator in the eigenbasis of the dipole interaction, with the energy scale defined by the dipole coupling constant $A$. The libary defines the matrix representation of the coupled system using the uncoupled basis $\left|S_1, S_2, m_1, m_2 \right\rangle$ and the definitions in the Sec. 3.5 of \cite{sakurai1995modern}. For details of the implementation, werefeer the reader to the User guide \cite{openMMFUserManual}.

This type of system of coupled system is present in the electronic ground states of alkali atoms \cite{PhysRevLett.114.240401} and single-charged atomic ions \cite{randall2015efficient}, which are currently the focus of intense experimental research in quantum technology.  The interaction between the nuclear and electronic magnetic dipole moments defines two sub-sets of states separated by an energy gap of order $A$, which has enabled the use of radio-frequency and microwaves for coherent control of quantum states  \cite{,randall2015efficient,PhysRevLett.114.240401}. The library let us to readily evaluate the time-evolution of the atomic states in such situations \cite{sinuco2019microwave}.


\section{\label{sec:Examples} Illustrative examples}

The main features of the library are illustrated with two examples, which are relevant across many experimental platforms. We present a selection of examples that illustrate the potential of OPENMMF for a wide range of applications. The example in  Sec. \ref{sec:DressedRabiOscillations} focuses on a two-level systems, which can be extended to study quantum control of quantum systems with a small number of consituents. In Section we study a modified

\subsection{\label{sec:DressedRabiOscillations} Rabi oscillations of a dressed qubit}
As our first example, we consider the time-evolution of a two-level system (or qubit) driven by two harmonic drives with the Hamiltonian:
\begin{eqnarray}
H &=& \hbar \omega_0 S_z + \hbar \Omega_{\textrm{X}} \cos(\omega_D t + \phi_D) S_x \nonumber \\ 
& & + \hbar \Omega_{P,x} \cos(\omega_P t + \phi_{P,x})S_x \nonumber \\
& &  + \hbar \Omega_{P,z} \cos(\omega_P t + \phi_{P,z})S_z
\label{eq:BichromaticTLS}
\end{eqnarray}
where the subindice $D$ ($P$) indicates that the mode is a Dressing (Probing) the qubit, in the sense defined in section \ref{sec:dressedbasis}. 

In this case, the data structures required to setup the matrix representation of the Hamiltonian components are declared with the instructions:

\begin{Verbatim}[fontsize=\scriptsize]
TYPE(ATOM)                 QUBIT
INTEGER,   DIMENSION(3) :: MODES_NUMBER
TYPE(MODE),DIMENSION(3) :: DRIVING_FIELD

MODES_NUMBER = (1,1,1)
\end{Verbatim}

The second and third components of the integer array \verb MODES_NUMBER   indicate that the qubit is driven by one harmonic of each frequency. The parameters of the drivings are defined via the components of the \verb MODE  data type \verb DRIVING_FIELD . Since a qubit is a spin $1/2$ system, an explicit declaration of the phase and amplitude of the driving fields is sufficient to define the Hamiltonian components. For driving mode $D$ this is done as follows:

\begin{Verbatim}[fontsize=\scriptsize]
DRIVING_FIELD(2)%X         = OMEGA_X
DRIVING_FIELD(2)%phi_x     = 0.0
DRIVING_FIELD(2)%omega     = OMEGA_D
\end{Verbatim}
with a similar set of instruction to define the parameters of the $P$ driving mode.

The calculation of the time-evolution operator in the bare basis is then done following the sequence of instructions detailed in Sec. \ref{sec:sequence}:
\begin{Verbatim}[fontsize=\scriptsize,commandchars=\\\{\},codes={\catcode`$=3}]
!===============================================================================
!== MULTIMODE TIME-EVOLUTION OPERATOR IN THE BARE BASIS
!===============================================================================
CALL SETHAMILTONIANCOMPONENTS(...,MODES_NUM,DRIVING_FIELD,INFO)
CALL MULTIMODEFLOQUETMATRIX(...,MODES_NUM,DRIVING_FIELD,INFO)          
CALL LAPACK_FULLEIGENVALUES(\textbf{U_F},SIZE(U_F,1),\textbf{E_FLOQUET},INFO)
!===============================================================================
!===== EVALUATE TIME-EVOLUTION OPERATOR BETWEEN T1 AND T2 IN THE BARE BASIS
!===============================================================================
CALL MULTIMODETIMEEVOLUTINOPERATOR(...,MODES_NUM,U_F,E_FLOQUET,...,DRIVING_FIELD, &
                                &   T1,T2,\textbf{U_BARE},INFO)
\end{Verbatim}
where the dots represent arguments omitted for shortness and clarity and the bold font indicates output subroutine arguments. This procedure results in the $2 \times 2$ matrix  \verb U_BARE  containing the time evolution from \verb T1  to  \verb T2 .

In this type of systems, it is of interest to calculate the time-evolution operator in the basis of dressed states defined by one of the modes, here selected as the $2$nd mode ($D$). The openMMF library let us define the Fourier decomposition of the required change of basis just by defining the subset of dressing modes as follows:

\begin{Verbatim}[fontsize=\scriptsize]
DRESSING_MODES = 2   ! including the static mode
DRESSING_FIELD_INDEX(1) = 1  ! The static mode 
DRESSING_FIELD_INDEX(2) = 2  ! The dressing mode 
CALL DRESSEDBASIS_SUBSET(...,DRESSING_FIELDS_INDEX,
        MODES_NUM,DRIVING_FIELD,U_FD,E_DRESSED,...) 
\end{Verbatim}
As a result of this sequence of instructions, we obtain a matrix with the Fourier decomposition of the micromotion operator associated with the subset of dressing modes, \verb U_FD  , along with an array containing the corresponding dressed eigenenergies \verb E_DRESSED . To calculate the time evolution in the dressed basis, \verb U_DRESSED, we should evaluate the micromotion operator at times \verb T_1 , \verb U_F_T1 ,  and \verb T_2 , \verb U_F_T2 , which are then used in the transformation Eq.~ (\ref{eq:baretimeevolution}). This is done with the following sequence of instructions:
	
\begin{Verbatim}[fontsize=\scriptsize,commandchars=\\\{\},codes={\catcode`$=3}]
CALL MULTIMODEMICROMOTION(...,U_FD,E_DRESSED,T_1,\textbf{U_F_T1},INFO)
CALL MULTIMODEMICROMOTION(...,U_FD,E_DRESSED,T_2,\textbf{U_F_T2},INFO)  
U_DRESSED = MATMUL(TRANSPOSE(CONJG(U_F_T2)),MATMUL(U_BARE,U_F_T1))
\end{Verbatim}
where the first and second lines produced the time-dependent micromotion operators of the dressed basis, and the third line executes the basis transformation. 

\begin{figure}[!htb]
\centering
\includegraphics[width=0.7\linewidth]{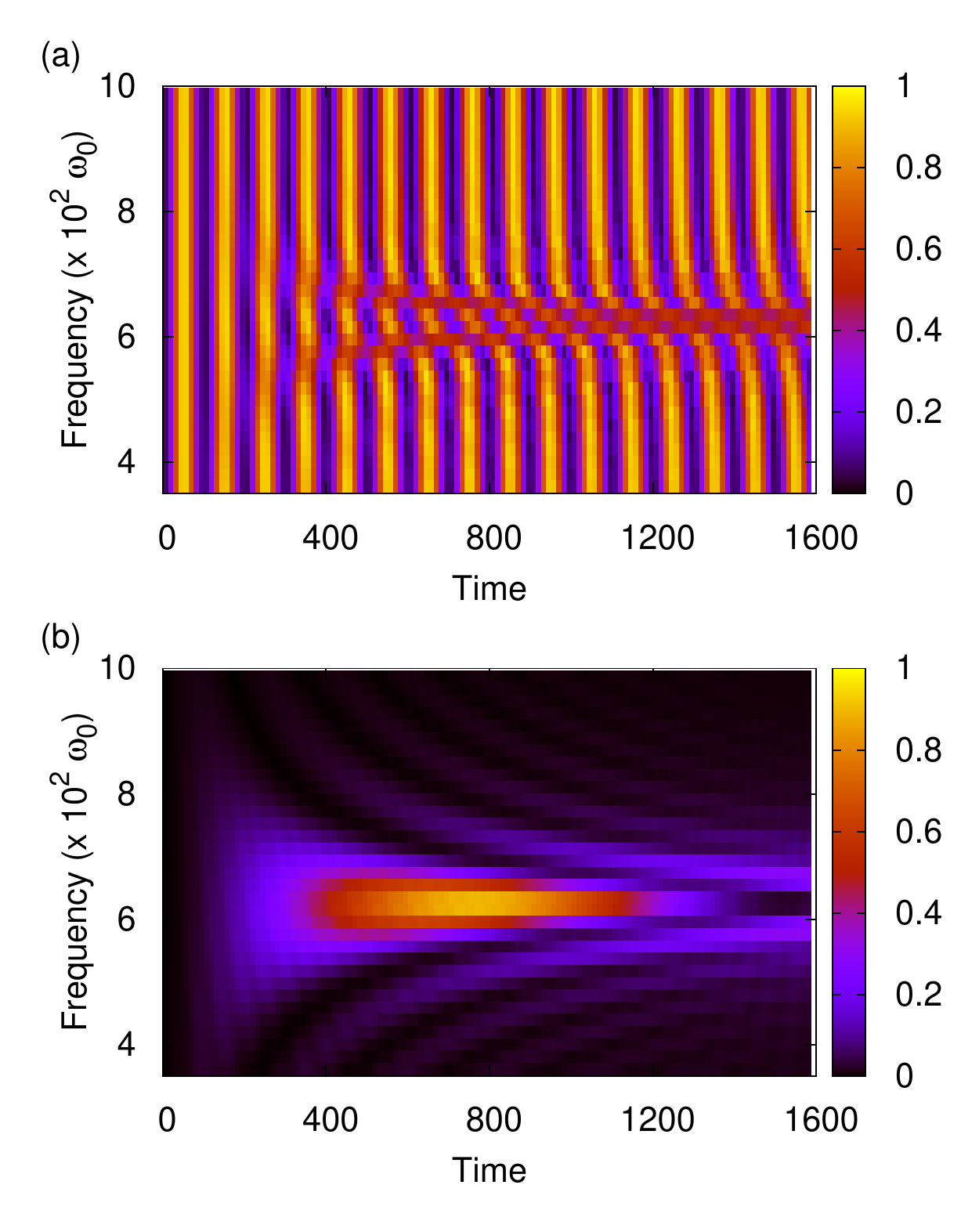}
\caption{\label{fig:qubitRabiOscillation} Probability of transition of a resonantly dressed qubit as a function of the time and probe frequency $\omega_P$ (a) in the basis of bare states (b) in the dressed basis defined. The Hamiltonian parameters are: $\hbar \omega_0 = 1.0, \hbar \omega_D = 0.025, \hbar \omega_D = 1.0, \hbar \Omega_{P,x} = 0.005,\hbar \Omega_{P,z} = 0.025$, with all phase factors set  to zero.}
\end{figure}

As a concrete example of the convinience of the dressed basis, in Fig.~\ref{fig:qubitRabiOscillation} we plot the time evolution of the probability of remaining in an initially pure state as a function of time and the frequency of the second driving field. Fast oscillations of this probability are displayed in the original qubit basis, while a slower dynamics is observed in the dressed basis. Note the resonant feature at the frequency $\omega_P = \Omega_{\textrm{x}}/2.0$ \cite{shirley1965solution}.

\subsection{\label{sec:oneDlatticeDriven} Harmonically driven 1-D lattice }
Technical advances in atomic and optical physics have enabled experiments with  ultracold atomic ensembles loaded in spatially periodic potentials landscapes defined by optical lattices \cite{goldman2016topological}. Controlled temporal modulation of the lattice parameters can induce non-equilibrium phenomena of interest for quantum simulations, and the experimental demonstration of such phenomena is leading to rapid advances in this area. Many important dynamical features of such driven systems are diagnosed via the dependence of eigenvalues and eigenvectors of the Hamiltonian in the frequency-domain with respect to the variation of the driving parameters \cite{desbuquois2017controlling,lellouch2017parametric}. 

To illustrate the use of the OPENMMF library for these problems, here we consider a spatially periodic one-dimensional lattice with a time-periodic modulation of the next-neighbour hopping. The lattice has \verb D/2  unit cells, each containing two sites with  energy offsets $\pm V_0$. The corresponding Hamiltonian reads:

\begin{eqnarray}
  H &=& \sum_{m=1}^D  (- 1)^m V_0 \left| m \right\rangle \left\langle m\right|  + \nonumber \\
    &&  \sum_{m=1}^{D-1} t e^{i \cos (2 \pi m /D)} \left| m \right\rangle \left\langle m+1\right|  +  \textrm{h.c.} + \nonumber \\
  && \sum_{m-1}^{D-1} V \cos(\omega t + \phi)) \left| m \right\rangle \left\langle m + 1 \right| + \textrm{h.c.}  \nonumber 
\label{eq:HamiltonianLattice}
\end{eqnarray}
where the first terms defines the energy offset at each lattice site, the second line defines a site-dependent tunneling rate between neighbouring sites and the last term defines a uniform (i.e. site-independent) modulation of the near-neighbour tunneling. 

\begin{figure*}
\centering
\includegraphics[width=\textwidth]{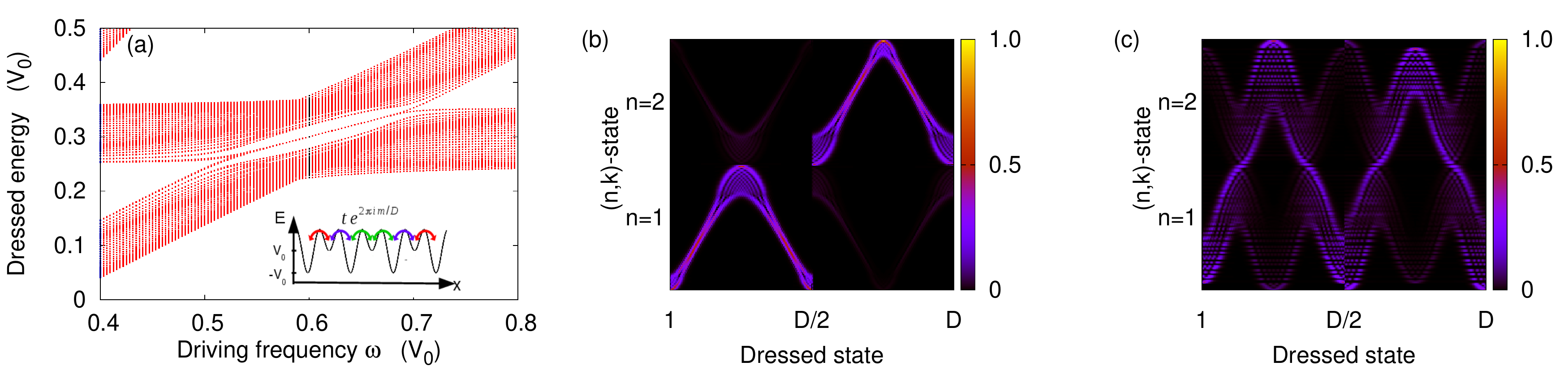}
\caption{\label{fig:shakenlattice}  Dressed spectrum of a driven lattice model. (a) Eigenenergies of the frequency-domain representation of the Hamiltonian Eq.~ (\ref{eq:HamiltonianLattice}) as a function of the driving frequency $\omega$. (inset) Periodic potential described by the tight-binding model Eq.~ (\ref{eq:HamiltonianLattice}). Projection of the dressed states onto the eigen-basis of the static lattice when the driving frequency  is (b)  red detuned ($\omega = 0.4$, black dots in panel (a)) (c) resonant with the band gap ($\omega = 0.6$, blue dots in panel (a)). The parameters of the model are: $D=256, V_0 = 0.25, t=0.125, V=0.0625$ and $\phi = 0$.}
\end{figure*}

The three Fortran data structures containing the parameters of the Hamiltonian are initialised as before, but here only two modes are needed (the static component plus one driving mode). The static and driving components of the Hamiltonian are declared with the following set of instructions:

\begin{Verbatim}[fontsize=\scriptsize]
DO m=1,D ! ALL  SITES
   DRIVING_FIELD(1)%V(m,m)  = ((-1)**m)*V_0
END DO

! NEAR NEIGHBOURS COUPLINGS
DO m=1,D-1
   !STATIC COMPONENT
   DRIVING_FIELD(1)%V(m,m+1)  = t*EXP(DCMPLX(0.0, 1.0)*COS(pi*2*m/D))
   DRIVING_FIELD(1)%V(m+1,m)  = t*EXP(DCMPLX(0.0,-1.0)*COS(pi*2*m/D))

   !DRIVING MODE
   DRIVING_FIELD(2)%V(m,m+1)  = V
   DRIVING_FIELD(2)%V(m+1,m)  = V
END DO
DRIVING_FIELD(2)%omega     = omega
\end{Verbatim}

\noindent
which displays a correspondence with the Hamiltonian Eq.~(\ref{eq:HamiltonianLattice}). Just two more instructions are sufficient to evaluate the eigenvalues and eigenenergies of the frequency-domain representation of the Hamiltonian:

\begin{Verbatim}[fontsize=\scriptsize]
CALL MULTIMODEFLOQUETMATRIX(..., MODES_NUMBER,DRIVING_FIELD,INFO)
CALL LAPACK_FULLEIGENVALUES(U_F,..., E_FLOQUET,INFO)
\end{Verbatim}

\noindent
The second line calls a wrapper to a diagonalisation subroutine from the LAPACK library. The resulting eigenvectors are stored in the matrix \verb U_F  and the eigenvalues in the array \verb E_FLOQUET . A typical example of this calculation is shown in Fig.~\ref{fig:shakenlattice}, where the set of eigenenergies are plotted as a function of the driving frequency. 

In addition to evaluation of the Floquet spectrum, the numerical calculation of the time-evolution operator is of interest for Hamiltonian engineering with atomic ensembles in atomic lattices \cite{Holthaus_2015}. This capability of the OPENMMF can be used to  evaluate corrections to analytical approximations of the Effective Hamiltonian \cite{Holthaus_2015}, and their effects on the on the dynamics of the atomc cloud \cite{PhysRevB.91.245135}.


\section{\label{impact} Impact}
Considering that the Hamiltonian in Eq.~ (\ref{eq:Hamiltonian}) describes a wide diversity of physical systems, the OPENMMF library constitutes a general-purpose tool for studying quantum dynamics of driven systems. The library is open-source and provides a ready-to-use interface, which can be exploited by users with a broad range of expertise ranging from graduate students to established researchers. An important feature of the library is the set of functions that allow the user to evaluate the time-evolution on a dressed basis, which is of current interest across different fields. To illustrate the degree of generality of the library, and therefore its potential impact, here we mention three specific research areas where the OPENMMF library can find the direct application.

\bigskip
\noindent
\textit{Coherent control of quantum systems}

Precise control over the state of a quantum system is key for quantum technologies. An important operational step in such applications is the precise creation of quantum superposition, which is generally achieved by interleaving periods of free and driven evolution described by a Hamiltonian of the form Eq.~ \ref{eq:Hamiltonian} \cite{macquarrie2015continuous}.  Decoherence induced by the interaction of the system with the environment is a major obstacle to achieve the levels of control required for the realisation of practical applications. However, superposition of dresses states can be more robust to environmental noise than bare states \cite{Mas_2019,SinucoMas2020}.  This mechanism to reduce the sensitivity of a quantum system to noise is known as Continuous Dynamical Decoupling and, to be implemented, it requires an experimental sequence that follows the path shown in Fig. \ref{fig:DressedBasis}. This technique has found application improving the coherence of atomic clocks \cite{sarkany2014controlling,PhysRevA.97.013407} and qubit gates \cite{randall2015efficient, PhysRevLett.113.237601}, as well as for the sensitivity and spatial resolution of low-frequency magnetometers operating \cite{yan2013rotating,baumgart2016ultrasensitive,hirose2012continuous}. 

\bigskip
\noindent
\textit{Time-domain quantum simulations}

The frequency-domain representation of the Hamiltonian Eq.~ (\ref{eq:Hamiltonian}) with $N$ mutually irrational frequencies is given by Eq.~ (\ref{eq:multimodeeigenproblem}). This representation can be interpreted as the  Hamiltonian of an $N$-dimensional lattice with sites label by the index $\vec{n}$,  a DC force equal to $\vec{\omega}$ and couplings between sites controlled by the driving modes. The typical signatures of dynamics of the simulated lattice (e.g. band structure, conductivity, edge-states, etc.) arise in the time-evolution of the system, which is directly accessible with the OPENMMF library. Topologically non-trivial states \cite{martin2017topological,ma2018experimental} and Anderson localisation in quasi-crystals \cite{delande2017three} are two phenomena accessible with this scheme for quantum simulations, which could be realised with a significantly simplified experimental setup.

\bigskip
\noindent
\textit{Out-of-equilibrium phases of driven interacting systems}

The behaviour of driven systems can be qualitatively very different from their static counterparts. Various driving schemes have been recently used to engineer effective Hamiltonian displaying out-of-equilibrium phases that are impossible to realise with static systems \cite{PhysRevB.82.235114}. Such applications are subject of intense experimental and theoretical research with ultracold atomic gases loaded in optical lattices \cite{Holthaus_2015}, where the Hamiltonian and driving schemes are similar to those in the example in Sec. \ref{sec:oneDlatticeDriven}. In the single-particle case, the dynamics in the limits of high frequency or strong and resonant driving are relatively well understood \cite{PhysRevB.91.245135}, and novel phases of matter in such regimes have been proposed and realised in recent years \cite{desbuquois2017controlling,nakajima2016topological}. However, there are several fundamental questions still to be addressed in the context of many-body driven systems, which emerge from the interplay between interactions, driving and symmetries/disorder \cite{delande2017three, PhysRevLett.116.250401}. The OPENMMF library provides a tool to investigate such phenomena, as we show in our last illustrative example. Though current computational limitations allows only to investigate systems with relatively small Hilbert space dimension, such studies provide a solid understanding of the non-equilibrium dynamics of driven systems \cite{PhysRevLett.119.123601,PhysRevLett.121.050602,PhysRevLett.121.076802} and the library provides a platform of simple usability.

\section{\label{impact} Conclusions}
In this paper, we introduce the OPENMMF library as a tool to calculate the time-evolution operator of quantum systems with discrete spectrum driven by harmonic couplings. The library implements the multimode Floquet expansion of a time-dependent Hamiltonian and provides several functionalities relevant for its analysis. In particular, the library allows the user to evaluate the micromotion operator associated with arbitrary drivings built as a superposition of harmonic terms. In passing, this procedure generalises the notion of dressed states to such situations. 

The general formulation implemented in the library enables its use for applications where determining the time-evolution operator is relevant, such as designing optimal control sequences, tailoring the response of quantum systems and studying non-equilibrium quantum dynamics and Hamiltonian engineering. The code is written in Fortran 90, and we provide wrappers for C++ and python, making it widely accessible and easy usage. 

\section*{Acknowledgements}
We acknowledge fruitful comments and input from Pedro Nevado Serrano, Juan Totero Gongora and Barry Garraway. This work has been supported by the EPSRC grants EP/I010394/1, EP/M013294/1 and the University of Sussex. Special acknoledgments to Camilo Guerrero and ConectLabs SAS (https://www.facebook.com/conectalabs/) for supporting code review of this project. 

\section*{Data Statement}
All relevant data presented in this work can be generated with the examples of the openmmf library (examples/FORTRAN, examples/CPP and examples/Python), which are available via the Elsevier/SoftwareX git repository (https://github.com/ElsevierSoftwareX/) and the OPENMMF project repository (https://github.com/openMMF/). The library source code is available as an open-source project through the same repositories, covered by a CC-BY 4.0 license. 

\section*{Appendix A: Typical example of the matrix representation of the multimode Hamiltonian}

A typical shape of the matrix representation of the right-hand-side of Eq.~ (\ref{eq:multimodeeigenproblem}). Here we consider a static system $H_0$ driven by the three frequencies: $\omega_1,2\omega_1,\omega_2$. The integer array describing this set of frequencies must be $(1,2,1)$. For the two fundamental frequencies, the number of Fourier modes in the decomposition of the micromotion operator is chosen to be $N_F=3$. With this, the driving of  $\omega_1$ and its first harmonic leads to the matrix:

\[
\mathcal{H}_1 = \begin{pmatrix}
 H_0 + 3 \hbar \omega_1 & V^{1,1} & V^{1,2} & 0 & 0 & 0 &0 \\
  V^{1,1\dagger} & H_0 + 2 \hbar  \omega_1 & V^{1,1} & V^{1,2} & 0 & 0 & 0 \\
 V^{1,2 \dagger} &  V^{1,1\dagger} & H_0 + \hbar \omega_1 & V^{1,1} & V^{1,2} & 0 & 0 \\
 0 &V^{1,2 \dagger} &  V^{1,1 \dagger} & H_0  & V^{1,1}  & V^{1,2} & 0  \\
 0 & 0 &V^{1,2 \dagger} &  V^{1,1 \dagger} & H_0 - \hbar \omega_1 & V^{1,1} & V^{1,2}   \\
 0 & 0 & 0 &V^{1,2 \dagger} &  V^{1,1 \dagger} & H_0 - 2 \hbar \omega_1 & V^{1,1} \\
 0 & 0 & 0 & 0 &V^{1,2 \dagger} &  V^{1,1 \dagger} & H_0 - 3 \hbar \omega_1 
\end{pmatrix}
\]
\noindent
and the full matrix results in:
\[
\mathcal{H} = \begin{pmatrix}
 \mathcal{H}_1 + 3 \hbar \omega_2 & \mathcal{V}^{2,1} & 0 & 0 & 0 & 0 &0 \\
  \mathcal{V}^{2,1\dagger} & \mathcal{H}_1 + 2 \hbar  \omega_2 & \mathcal{V}^{2,1} & 0 & 0 & 0 & 0 \\
 0 &  \mathcal{V}^{2,1\dagger} & \mathcal{H}_1 + \hbar \omega_2 & \mathcal{V}^{2,1} & 0 & 0 & 0 \\
 0 & 0 &  \mathcal{V}^{2,1 \dagger} & \mathcal{H}_1  & \mathcal{V}^{2,1}  &0 & 0  \\
 0 & 0 & 0 &  \mathcal{V}^{2,1 \dagger} & \mathcal{H}_1 - \hbar \omega_2 & \mathcal{V}^{2,1} & 0   \\
 0 & 0 & 0 &0 &  \mathcal{V}^{2,1 \dagger} & \mathcal{H}_1 - 2 \hbar \omega_2 & \mathcal{V}^{2,1} \\
 0 & 0 & 0 & 0 & 0 &  \mathcal{V}^{2,1 \dagger} & \mathcal{H}_1 - 3 \hbar \omega_2
\end{pmatrix}
\]
with
\[
\mathcal{V}^{2,1} = \begin{pmatrix}
  V^{2,1} & 0 & 0 & 0 & 0 & 0 &0 \\
  0&V^{2,1} & 0 & 0 & 0 & 0 & 0  \\
  0&0&V^{2,1} & 0 & 0 & 0 & 0  \\
  0&0&0&V^{2,1} & 0 & 0 & 0 \\
  0&0&0&0&V^{2,1} & 0 & 0  \\
  0&0&0&0&0&V^{2,1} & 0 \\
  0&0&0&0&0&0&V^{2,1}  
\end{pmatrix}
\]

\section*{C++ wrapper}
The library includes wrappers to use with C++. The interfaces to the Fortran subroutines  share the name of the aimed function with the appended particle \verb _C_ .  The full set of wrappers are declared in the file \verb MultimodeFloquet.h  , which must be included in the C++ source code. As an example, we present a C++ program that evaluates the time-evolution of a driven qubit.

\begin{verbatim}

#include <iostream>
#include <complex>
#include <stdio.h>
#include <math.h>
#include <string.h>

using namespace std;
typedef std::complex<double> dcmplx;

#include "MultimodeFloquet.h"

extern "C" int h_floquet_size;

int main(){

  atom_c id;
  int info,N_;
  int jtotal;
  char name[]     = "qubit";
  char manifold[] = "U";

  int r,m,l;
  int d_bare,total_frequencies;

  double t1,t2;

  info   = 0;
  jtotal = 2;
  floquetinit_c(&id,name,&info);

  d_bare = id.d_bare;

  dcmplx * U_AUX = new dcmplx [d_bare*d_bare];

  int nm = 2;
  int * modes_num = new int [nm];

  modes_num[0] = 1;
  modes_num[1] = 1;
  
  total_frequencies = 0;
  for(r=0;r<nm;r++){
    total_frequencies += modes_num[r];
  }
  
  mode_c_f * fields = new mode_c_f [total_frequencies];

  // ALLOCATE MEMORY FOR THE COUPLING MATRICES
  for(r=0;r<total_frequencies;r++){
    fields[r].V = new dcmplx *[d_bare];
    for(l=0;l<d_bare;l++){
      fields[r].V[l] = new dcmplx [d_bare];
    }
  }
  

  // --- SET DRIVING PARAMETERS   
  fields[0].x    = 0.0;
  fields[0].y    = 0.0;
  fields[0].z    = 2.0;
  fields[0].phi_x = 0.0;
  fields[0].phi_y = 0.0;
  fields[0].phi_z = 0.0;
  fields[0].omega = 0.0;
  fields[0].N_Floquet = 0;

  fields[1].x     = 4.0;
  fields[1].y     = 0.0;
  fields[1].z     = 0.0;
  fields[1].phi_x = 0.0;
  fields[1].phi_y = 0.0;
  fields[1].phi_z = 0.0;
  fields[1].omega = 2.0;
  fields[1].N_Floquet = 3;

  // --- SET THE HAMILTONIAN COMPONENTS
  coupling_init(fields,&total_frequencies,&d_bare,&info);
  sethamiltoniancomponents_c_
               (&id,&nm,&total_frequencies,modes_num,&info);
    
  // --- BUILD THE MULTIMODE FLOQUET MATRIX AND  FIND ITS SPECTRUM     
  multimodefloquetmatrix_c_
   (&id,&nm,&total_frequencies,modes_num,fields,&info);

  double * e_floquet = new double [h_floquet_size];
  dcmplx * U_F =  new dcmplx [h_floquet_size*h_floquet_size];
    
  lapack_fulleigenvalues_c_
   (U_F,&h_floquet_size,e_floquet,&info);
// the diagonalization is done with the internal (Fortran)
// Hamiltonian (H_FLOQUET) and the calculated
// U_F is the transformation that diagonalise this Hamiltonian
// On the Fortran side, H_FLOQUET is deallocated after
// diagonalization. 

   //--- EVALUATE TIME-EVOLUTION OPERATOR IN THE BARE BASIS

   N_ = 256;
   t1= 0.0;
   for(r=1;r<N_;r++){      
     t2 = r*100.0/N_;
     multimodetimeevolutionoperator_c_
       (&h_floquet_size,&nm,modes_num,U_F,e_floquet,&d_bare,
        &t1,&t2,U_AUX,&info);
     for(l=0;l<d_bare*d_bare;l++) p_avg[l] = pow(abs(U_AUX[l]),2);
     write_matrix_c_(p_avg,&d_bare);           
   }
    delete(e_floquet);    
    delete(U_F);
  }
  
  return 0;
}

\end{verbatim}

\section*{Python wrapper}
As an experimental feature, the library includes wrappers to use with Python through calls to the C++ wrapper functions using \verb ctypes .  The full set of wrapper functions are declared in the file \verb src/openmmf.py , which must be included in working directory of the python script.  As an example, we present a Python script that evaluates the time-evolution of a driven qubit. Other examples can be found in the folder \verb examples/PYTHON/ .

\begin{verbatim}
import numpy as np
import openmmf as openmmf
import matplotlib.pyplot as plt

# INITIALISE THE DATA TYPE
id  = openmmf.atom_c_T()
name         = 'qubit'
info = 0

openmmf.floquetinit(id,name,info=info)
d_bare = id.d_bare

# DEFINE THE NUMBER OF MODES
modes_num = np.array([1,1],dtype=np.int32)
nm        = np.sum(modes_num)

# THIS INSTRUCTION DEFINES A TYPE OF 
# ARRAY OF modes WITH nm COMPONENTS
fields    = openmmf.mode_c_T*nm 

# THIS INSTANCE DECLARES THE FIELDS
field     = fields()

# DEFINE EACH ONE OF THE DRIVING FIELDS
field[0].x = 0,0
field[0].y = 0,0
field[0].z = 1,0
field[0].phi_x = 0
field[0].phi_y = 0
field[0].phi_z = 0
field[0].omega = 0
field[0].N_Floquet = 0

field[1].x = 2,0
field[1].y = 0,0
field[1].z = 0,0
field[1].phi_x = 0.0
field[1].phi_y = 0.0
field[1].phi_z = 0.0
field[1].omega = 1.0
field[1].N_Floquet = 8

N_= 128
M_= 128

# declare arrays to store the solution
P_AVG      = np.zeros([N_,d_bare*d_bare],dtype=np.double)
P_TimeEvol = np.zeros([N_,N_,d_bare*d_bare],dtype=np.double)

for m in range(N_):
    # SET A NEW FIELD CONFIGURATOIN, E.G.
    field[1].omega = 0.2 + m*2.0/N_

    # SET THE HAMILTONIAN COMPONENTS    
    openmmf.sethamiltoniancomponents(id,modes_num,field,info)

    # BUILD THE MULTIMODE FLOQUET MATRIX
    h_floquet_size = openmmf.multimodefloquetmatrix(id,modes_num,field,info)

    # ALLOCATE ARRAYS FOR FLOQUET ENERGIES (E_FLOQUET), MICROMOTION (U_F), 
    # TIME-AVERAGE TRANSITION PROBABILITY (p_avg), AND 
    # AN AUXILIAR MATRIX (U_AUX)
    e_floquet = np.zeros(h_floquet_size,dtype=np.double)
    U_F       = np.zeros(h_floquet_size*h_floquet_size,dtype=np.complex)
    p_avg     = np.zeros(id.d_bare*id.d_bare,dtype=np.double)
    U_AUX     = np.zeros(id.d_bare*id.d_bare,dtype=np.complex)

    # DIAGONALISE THE MULTIMODE FLOQUET MATRIX
    openmmf.lapack_fulleigenvalues(U_F,h_floquet_size,e_floquet,info)

    # EVALUATE THE AVERAGE TRANSITION PROBATILIBIES IN THE BARE BASIS
    openmmf.multimodetransitionavg(h_floquet_size,field,modes_num,
    U_F,e_floquet,d_bare,p_avg,info)
    P_AVG[m,:] = p_avg

    #SET THE INITIAL (T1) AND FINAL TIME (T2)
    t1     = 0.0
    for r in range(M_):
        t2 = r*32.0*4.0*np.arctan(1.0)/M_ 
        # EVALUATE THE TIME EVOLUTION OPERATOR
        openmmf.multimodetimeevolutionoperator(h_floquet_size,modes_num,U_F,e_floquet,d_bare,field,t1,t2,U_AUX,info)
        P_TimeEvol[m,r,:] = np.power(np.abs(U_AUX),2)

# DEALLOCATE ALL ARRAYS
# WARNING: THIS FUNCTION SHOULD BE CALLED TO CLEAR MEMORY 
openmmf.deallocateall(id)

# PLOTTING
omega      = np.linspace(0.2,2.2,N_)
#plotting the avg transition probability
plt.plot(omega,P_AVG[:,1])
plt.xlabel('frequency')
plt.ylabel('Average transition probability')
plt.show()

#%%
# plot the time evolution
t = np.linspace(0,32.0*4.0*np.arctan(1.0),M_)
X,Y = np.meshgrid(t,omega)
Z = P_TimeEvol[:,:,1]

fig,ax = plt.subplots()
im = ax.pcolormesh(Y, X, Z,shading='auto')
ax.set_ylabel('time')
ax.set_xlabel('frequency')
fig.tight_layout()

\end{verbatim}

\bibliography{LibraryBib}

\end{document}